\documentclass[10pt,twocolumn,journal]{IEEEtran}

\IEEEoverridecommandlockouts

\usepackage{srcltx} 
\usepackage{amsmath,amsxtra,amssymb,amsthm}
\usepackage{times,latexsym,array,verbatim,url}
\usepackage{color,graphicx,cite}
\usepackage{datetime}

\newcommand{\techabbr}{Xtreaming}

\title{Xtreaming: An Incremental Multidimensional Projection Technique and its Application to Streaming Data}

\author{
    \IEEEauthorblockN{
        T\'acito Tiburtino\IEEEauthorrefmark{1},
        Rafael M. Martins\IEEEauthorrefmark{2},
        Danilo B. Coimbra\IEEEauthorrefmark{3},\\
        Kostiantyn Kucher\IEEEauthorrefmark{2},
        Andreas Kerren\IEEEauthorrefmark{2},
        Fernando V. Paulovich\IEEEauthorrefmark{4}
    }\\
    \vspace{1em}
    
    \IEEEauthorblockA{
    \IEEEauthorrefmark{1}\textit{PAIVA - NCEx} - 
    \textit{Federal University of Alagoas (UFAL)}, Arapiraca, AL, Brazil \\
    \texttt{tacito.neves@arapiraca.ufal.br}
    }
    \vspace{.25em}
     
    \IEEEauthorblockA{
    \IEEEauthorrefmark{2}\textit{Department of Computer Science and Media Technology} - \textit{Linnaeus University}, V\"axj\"o, Sweden \\
    \texttt{\{rafael.martins,kostiantyn.kucher,andreas.kerren\}@lnu.se}
    }
    \vspace{.25em}
     
    \IEEEauthorblockA{
    \IEEEauthorrefmark{3}\textit{Computer Science Department} - \textit{Federal University of Bahia (UFBA)}, Salvador, BA, Brazil \\
    \texttt{coimbra.danilo@ufba.br}
    }
    \vspace{.25em}
     
    \IEEEauthorblockA{
    \IEEEauthorrefmark{4}\textit{Faculty of Computer Science, Dalhousie University}, Halifax, Nova Scotia, Canada\\
    \texttt{paulovich@dal.ca}
    }
}

\date{March 2020}

\begin{document}

\maketitle
\thispagestyle{empty}
\pagestyle{empty}


\begin{abstract}
Streaming data applications are becoming more common due to the ability of different information sources to continuously capture or produce data, such as sensors and social media. Despite recent advances, most visualization approaches, in particular, multidimensional projection or dimensionality reduction techniques, cannot be directly applied in such scenarios due to the transient nature of streaming data. Currently, only a few methods address this limitation using online or incremental strategies, continuously processing data, and updating the visualization. Despite their relative success, most of them impose the need for storing and accessing the data multiple times, not being appropriate for streaming where data continuously grow. Others do not impose such requirements but are not capable of updating the position of the data already projected, potentially resulting in visual artifacts. In this paper, we present \textit\techabbr, a novel incremental projection technique that continuously updates the visual representation to reflect new emerging structures or patterns without visiting the multidimensional data more than once. Our tests show that \textit\techabbr~is competitive in terms of global distance preservation if compared to other streaming and incremental techniques, but it is orders of magnitude faster. To the best of our knowledge, it is the first methodology that is capable of evolving a projection to faithfully represent new emerging structures without the need to store all data, providing reliable results for efficiently and effectively projecting streaming data.
\end{abstract}


\section{Introduction} \label{sec:intro}


Prompted by the massive quantity of available data, our society is facing a paradigm shift towards making decisions based on more data-driven processes~\cite{butler2006nature}. Although a positive trend, it presents several challenges for the existing visualization techniques, from visual and computational scalability to latency issues. One particular type of scenario that presents requirements difficult to fulfill involves streaming data. Compared to the more conventional static applications where data is first recorded in persistent tables to be later processed, in streaming applications the data needs to be processed as it is produced or received since it is typically discarded after that~\cite{Gama2012streaming}. Examples of such applications include environmental monitoring~\cite{5471680}, network intrusion detection~\cite{WANG200858}, and real-time social media analysis~\cite{6152108}.

Among the existing visualization techniques devoted to processing significant amounts of data, multidimensional projection or dimensionality reduction techniques~\cite{8383983} are emerging as a fundamental tool by allowing the identification and analysis of similarity and neighborhood relationships and patterns~\cite{Joia2011}. Although many projection techniques can successfully handle large datasets, most are not appropriate for streaming data given its transient nature. Currently, only a few methods address this limitation by using online or incremental strategies, continuously processing data, and updating the visualization. 

Despite the success of such methods, they present significant bottlenecks that impair their practical usage. Some methods impose the storage and multiple accesses to all (received) data, being appropriate to incremental and progressive scenarios but not to streaming where data continuously grow. Others create one projection function at the beginning of the process and use it for all data, in general, resulting in low quality layouts since the function is not adapted to new structures or patterns that emerge over time. Finally, a few methods evolve the projection function to capture new incoming structures and patterns. However, the already projected instances are not updated to comply with the new function, potentially resulting in visual artifacts produced by overlapping subsequent misaligned projections. 

In this paper, we present \textit\techabbr, a novel incremental projection technique to address the challenging design conflict of updating over time the position of the already projected instances without storing or revisiting all data that has been already processed. \textit\techabbr~combines a change detection approach, an out-of-sample projection technique, and a novel re-projection strategy to update the projection without revisiting the multidimensional input data. \textit\techabbr~is competitive in terms of global distance preservation if compared to other streaming and incremental techniques, but it is orders of magnitude faster. In summary, the main contributions of this paper are:

\begin{itemize}
    \item A precise and fast multidimensional projection technique, called \textit\techabbr,~that can be efficiently and effectively used in streaming scenarios where data needs to be processed as received;
    \item A novel strategy to re-project the already processed data to comply with a new projection function without the need for revisiting the original multidimensional data, a critical aspect of any technique designed to handle streaming data that is usually discarded after being processing.
\end{itemize}

To the best of our knowledge, this is the first time a multidimensional projection or dimensionality reduction technique is capable of evolving a projection over time to faithfully represent new emerging structures without the need of traversing the multidimensional data more than once.

\section{Related Work}\label{sec:related}

Multidimensional projection techniques create computational models that map data instances into graphical elements preserving in the visual representation the pairwise distances calculated among the instances~\cite{8383983}. One existing classification splits them into two groups, the \textit{in-sample} and the \textit{out-of-sample} techniques~\cite{bengio2003outofsample}. While in-sample techniques produce layouts processing the data instances altogether, the out-of-sample ones initially select and project a small sample of the dataset and then map the remaining instances interpolating the sample projection. 

Regarding the overall distance and neighborhood preservation, in-sample techniques usually produce more precise results but incur on high computational costs. Some examples include Glimmer~\cite{ingram2009glimmer}, t-SNE~\cite{van2008visualizing}, classical Multidimensional Scaling (MDS)~\cite{torgerson1965multidimensional}, Sammon's Mapping ~\cite{sammon1969nonlinear}, and Local Linear Embedding (LLE)~\cite{roweis2000nonlinear}. In contrast, out-of-sample techniques are capable of handling much larger datasets in a fraction of the running time, but with a penalty in precision. In fact, most out-of-sample techniques are approximations of in-sample techniques where the in-sample strategies are used to project the initial sample. Pekalska et al.~\cite{pekalska1999new} proposes an approximation of Sammon's Mapping. Landmarks MDS (L-MDS)~\cite{de2004sparse} and Pivot MDS~\cite{brandes2007eigensolver} are approximations of MDS. Landmarks ISOMAP (L-ISOMAP)~\cite{silva2002global} and Landmarks LLE (L-LLE)~\cite{bengio2003outofsample} are based on ISOMAP and LLE, respectively. 

Recently, some out-of-sample techniques have been developed not as approximations of in-sample methods, but as strategies to allow the incorporation of user knowledge in the projection process. Examples include Least Squares Projection (LSP)~\cite{paulovich2008least}, Piecewise-Laplacian Projection (PLP)~\cite{paulovich2011piece}, and Local Affine Multidimensional Projection (LAMP)~\cite{Joia2011}. Although out-of-sample techniques can handle large datasets, they are not appropriate for scenarios where (new) data is continuously fed into the process. Since the quality of the out-of-sample techniques typically depends on the quality of the initial sample, all data has to be known before the process starts to ensure that the sampling process collects instances that faithfully represent the data distribution. To tackle this limitation, online strategies have been devised, processing data as received.

Basalaj~\cite{basalaj1999Incremental} presents an online version of MDS. In this technique, when a new instance is received, MDS is applied considering the new instance and the already processed ones, creating a new full pairwise distance matrix. Similarly, Alsakran~\cite{alsakran2011streamit} employs a force-based approach that is updated to consider new instances, also recomputing an in-memory full pairwise distance matrix. Jenkins et al.~\cite{jenkins2004spatio} and Law et al.~\cite{law2004nonlinear} present online versions of ISOMAP. In both cases, the techniques are focused on updating the (geodesic) distances. When a new instance is received, all instances are processed, and a full pairwise distance matrix is necessary. Law et al.~\cite{law2006incremental} speed-up this process by avoiding the need for a full pairwise matrix by presenting an online version of L-MDS. In this version, only the distances between the new instance to all other instances are needed. Kouropteva et al.~\cite{kouropteva2005incremental} and Schuon et al.~\cite{schuon2008truly} present online versions of LLE technique by defining strategies to update the neighborhood relationships when a new data instance is received.
Common to all these online approaches is the need for storing and revisiting all data during the projection process. Consequently, they are not appropriate for streaming scenarios where data continuously grow.

Different methods address this storage problem. Partial-Linear Multidimensional Projection (PLMP)~\cite{paulovich2010plmp} defines a strategy that artificially generates an initial sample without visiting the data, using this sample to create a single projection function employed to process all data. Saul et al.~\cite{saul2003think} devise a streaming version of LLE that creates a projection function considering the first incoming instances and employs this to process all data. Using a similar strategy, Mahapatra and Chandola~\cite{mahapatra2017s} propose a streaming adaptation of the ISOMAP technique. Although such methods can handle streaming data, creating one projection function at the beginning of the process and using it to project all data is a bottleneck. The computed projections will only be satisfactory if the new incoming data is similar to the data employed to create the initial function. If new structures or patterns emerge over time, these techniques are not able to represent them since the function does not evolve with the data. Thereby, in general, the produced layouts are of poor quality.

Towards the challenging idea of evolving the projection function over time, Len et al.~\cite{leng2014locally} present a streaming version of LLE. This technique stores the necessary parameters to construct the projection function and adapts them to consider the new incoming data. Similarly, Ross et al.~\cite{Ross2008} store the parameters of a PCA function, updating the PCA eigenbasis using the Sequential Karhunen-Loeve algorithm~\cite{levey2000skl} as instances are received. Fujiwara et al.~\cite{fujiwara2019incremental} uses the same strategy to evolve PCA projections but also use Procrustes analysis~\cite{gower2004procrustes} to align consecutive projections to preserve the user's mental map. Supervised strategies have also been developed to evolve with the data, such as the online versions of Maximum Margin Criterion~\cite{yan2006scalable} and Linear Discriminant Analysis~\cite{ye2005idr}. In all these cases, although the projection function changes to capture new incoming structures and patterns, the already projected instances are not adapted to the new function. Consequently, there could exist a misalignment between the placement of the already projected instances and the position of the new instances. This misalignment is a significant bottleneck for visualization purposes since meaningless visual patterns can be obtained by overlapping subsequent (misaligned) projections. 

In this paper, we present \textit\techabbr, a novel multidimensional projection technique for mapping streaming data. \textit\techabbr~addresses the limitations mentioned above through a framework capable of evolving a projection over time to represent new emerging dissimilarity structures while adapts the previously projected data without revisiting the multidimensional data already processed. It is detailed in the next section.


\section{Streaming Multidimensional Projection}

\subsection{Problem Formalization}

To set notation, let $X=\{x_1,x_2,\ldots,x_n\} \in \mathbb{R}^m$ be a multidimensional dataset with $\delta(x_i,x_j)$ a dissimilarity function between two data instances, and $Y=\{y_1,y_2,\ldots,y_n\} \in \mathbb{R}^2$ its mapping to the visual space with $d(y_i, y_j)$ a distance function between two graphical elements. A batch (in-sample or out-of-sample) multidimensional projection technique is a bijective function $f: X \rightarrow Y$ that maps $X$ into $Y$ preserving the dissimilarity or neighborhood structure of multidimensional datasets that are fully stored in persistent tables.

In streaming scenarios, the dataset $X$ is, however, not entirely available at the beginning of the process, and $k $ data partitions ${X=X_1 \cup X_2 \cup \ldots \cup X_k}$, with $X_i \cap X_j= \emptyset, \forall~X_i, X_j$ with $i \neq j$, are successively received. Considering a progressive scenario where data is projected as obtained and then discarded, it is not possible to have one function to project all the data, instead it is necessary a set of $r$ functions to project the first $r$ partitions ($r < k$) so that the union of the partitions' projections approximates the projection of all received data, that is, 
\begin{equation}
Y_{[1 \ldots r]} = f^1(X_1) \cup f^2(X_2) \cup \ldots \cup f^r(X_r) \approx f(X_1 \cup X_2 \cup \ldots \cup X_r)
\end{equation}
where $f^i$ represents the function employed to project the partition $X_i$. 

Although the state-of-the-art incremental/online projection techniques are based on such approximation, if new structures emerge over time, such as groups of similar instances, this can fail to differentiate them from the existing structures. Therefore, for an effective streaming projection strategy, not only the projection function needs to be updated to capture new structures, but also the projection ${Y_{[1 \ldots r-1]} = Y_1 \cup Y_2 \cup \ldots \cup Y_{[r-1]}}$ of the previous partitions needs to be adapted to comply with the new projection function $f^r$, that is 
\begin{equation}
Y_{[1 \ldots r]} = f^r(X_1) \cup f^r(X_2) \cup \ldots \cup f^r(X_{r-1}) \cup f^r(X_r).
\end{equation}

However, in data streaming scenarios, where storing all data is unfeasible, most of the partitions $X_1, X_2, \ldots, X_{r-1}$ are no longer available when projecting $X_r$, so ${f^r(X_1) \cup f^r(X_2) \cup \ldots \cup f^r(X_{r-1})}$ cannot be computed.

\vspace{-1em}
\subsection{Overview}

Our approach, \textit\techabbr, addresses both challenges, namely, (1) to update the projection function to represent new emerging dissimilarity structures, and (2) to adapt the current projection to comply with the updated function without revisiting the multidimensional data already processed. \textit\techabbr~handles the former by using a change detection strategy to trigger the reconstruction of a new function every time a change in data distribution is detected, and the latter by using the information contained in the current projection to evolve the visual representation over time. Notice that, instead of directly processing the data as received in partitions,  we use buffers of fixed sizes so that our approach is independent of partitions' sizes, and buffering strategies can be used to process the data in constant rates without depending on the speed the data is being received. In the rest of this text, the terms partition and buffer are interchangeably used. 

\begin{figure}[htb]
\centering
\includegraphics[width=\columnwidth]{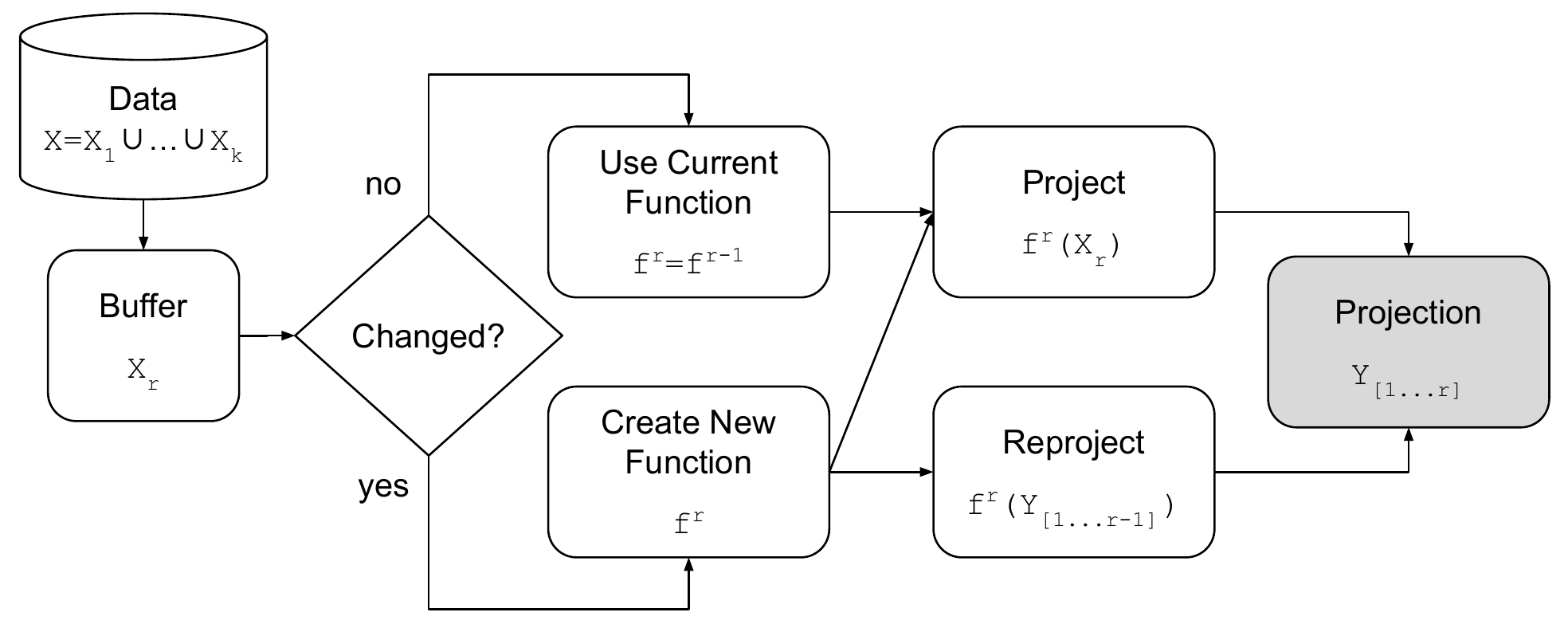}
\caption{\textit\techabbr~overview. Every time a new data buffer is received, we check if it represents a changing in the data distribution. If no changes are detected, the buffer is projected using the current projection function. Otherwise, the projection function is updated, the already projected data is re-projected, and the buffer is finally projected.}
\label{fig:overview}
\end{figure}

Figure~\ref{fig:overview} depicts the outline of our technique. The first buffer is used to initialize the process, resulting in the first projection function ($f^1$). After that, if a new incoming buffer (partition) does not represent a change in the distribution, it is projected using the current projection function. If a change is detected, the projection function is recreated, the data already projected is re-projected to comply with the new function, and finally, the buffer is projected. This process is repeated while there is incoming data. The next sections detail each step involved in our approach.
 
\vspace{-1em}
\subsection{Constructing the Projection Function}\label{sec:projfunc}

As previously discussed, if an incoming buffer $X_r$ does not represent a changing in distribution (we discuss that in Section~\ref{sec:sampling}),  the current projection function $f^{r-1}$ is used to process it, that is, $f^r=f^{r-1}$. Consequently, such a function needs to be ``stored'' or represented without accessing all the data.  To address this issue, we use an out-of-sample strategy. Out-of-sample strategies are two-step approaches which firstly project a small sample $\overline{X}$ of the dataset, with $|X| \gg |\overline{X}|$, and then interpolate the remaining instances using this initial projection as a base. Hence, we only need to store a small portion of the dataset (and a few parameters of the technique) to represent a projection function, obeying the data storage constraints for streaming scenarios. 

Besides being out-of-sample, the projection technique needs to fulfill other requirements: (1) it has to be fast and precise so that the running time and the quality of the produced mappings are not impaired; and (2) it needs to be a distance-based strategy (receive distances as input) so that the re-projection of the projected data can be performed without accessing all data (detailed in the next section). Currently, only a few out-of-sample techniques can comply with such requirements, such as the Pekalska approximation~\cite{pekalska1999new}, L-MDS~\cite{de2004sparse}, and UPDis~\cite{neves2018updis}. In this paper, we use the UPDis technique since it renders better results regarding distance preservation (see~\cite{neves2018updis} for comparisons) while presents competitive running times.

\vspace{-1em}
\subsection{Re-Projection}\label{sec:reproject}

Every time a new buffer $X_r$ is projected, the projection $Y_{[1 \ldots r-1]}$ of the previous buffers $X_{[1 \ldots r-1]}= X_1 \cup X_2 \cup \ldots \cup X_{r-1}$ needs to be updated to comply with the new projection function $f^r$. Given that we are using a distance-based out-of-sample technique, for re-projecting $X_{[1 \ldots r-1]}$ it is only necessary to compute the distances between $x_i \in X_{[1 \ldots r-1]}$ and the samples $\overline{x}_j \in \overline{X}_{[1 \ldots r]}$, where $\overline{X}_{[1 \ldots r]}$ represents the current sample, including the new instances detected by the change detection method (discussed in Section~\ref{sec:sampling}). 

The problem is, since $X_{[1 \ldots r-1]}$ is not stored, the distances among them and the samples cannot be calculated. We address such a limitation using a simple but effective strategy. Considering that the projection function is precise in terms of distance preservation (first requirement of the previous section), the distances between the instances $X_{[1 \ldots r-1]}$ and the samples $\overline{X}_{[1 \ldots r-1]}$ can be approximated, with some degree of accuracy,  by replacing the original distances by the projected distances, that is, by taking
\begin{equation}
 \delta(x_i,\overline{x}_j) \approx d(y_i,\overline{y}_j), \forall~x_i \in X_{[1 \ldots r-1]}, \overline{x}_j \in  \overline{X}_{[1 \ldots r-1]}
\end{equation}
and this information can be fully recovered from $Y_{[1 \ldots r-1]}$. 

For a complete projection, the only information that is missing is the distance between the samples $\overline{X}_r$ discovered by the change detection method and the instances already projected. We handle this problem by projecting $\overline{X}_r$ using the previous projection function $f^{r-1}$, using this projection to recover the distances, that is, by taking
\begin{equation}
 \delta(x_i,\overline{x}_j) \approx d(y_i, f^{r-1}(\overline{x}_j)), \forall~x_i \in X_{[1 \ldots r-1]}, \overline{x}_j \in \overline{X}_r
\end{equation}

Since $|X_r| \gg |\overline{X}_r|$, this process does not affect the running time of our approach. Notice that, since a high-precision out-of-sample strategy is employed, this is a good approximation for global distance preservation, and only uses the current projection and the sample, not requiring access to the instances' coordinates in $\mathbb{R}^m$ of the projected data. 

\vspace{-1em}
\subsection{Change Detection and Sampling}\label{sec:sampling}

Without loss of generality, we have presented our approach as if the projection function was updated every time a new buffer is received. Indeed, it is only updated if the incoming buffer contains new information, in terms of data distribution, when compared to the already processed buffers. Here we combine a clustering technique with an outlier detection strategy composing a two-step approach to detect changes on data streams.

In the first step, when a buffer $X_r$ is received, some representative samples are recovered to represent the buffer distribution. To do this, a distance-based clustering technique, called bisecting k-means~\cite{steinbach2000comparison}, is applied splitting $X_r$ into $q$ disjoint clusters $X_r=\overline{C}_1 \cup \ldots \cup \overline{C}_q$ containing similar instances. Then, for each cluster $\overline{C}_i$, its medoid $\overline{x}^m_i$ (the instance closest to its centroid) is selected as a representative, composing a pre-sample $\overline{X}^m = \overline{x}^m_i, \ldots, \overline{x}^m_q$. Based on a common heuristic, we set the number of clusters to $q=\sqrt{|X_r|}$ since it defines a good upper-bound for the number of clusters~\cite{paulovich2011piece, paulovich2010plmp, Joia2011}. Other sampling methods can also be used in this process, from simple random selections to strategies based in spectral decomposition~\cite{joia2015uncovering}. Here we use a cluster-based strategy given the good tradeoff in terms of running time and quality of the recovered samples. 

In the second step, each medoid $\overline{x}^m_i$ in the pre-sample $\overline{X}^m$ is tested to verify if it should be added to the partition sample $\overline{X}_r$ or not. The reason is to avoid that instances already represented in the current sample $\overline{X}_{[1 \dots r-1]}$ be added to the buffer sample, making $\overline{X}_r$ contain only new information. In this process, we use the Incremental Local Outlier Factor (iLOF)~\cite{pokrajac2007incremental} technique to measure how different a medoid is from the current sample using the concept of local densities. In this process, for each medoid $\overline{x}^m_i$ we first calculate its nearest neighbors considering the current sample $\overline{X}_{[1 \dots r-1]}$. If the medoid lies in a region of the multidimensional space with substantially lower density than its nearest neighbors, it is added to the sample $\overline{X}_r$. Although a precise and well-established method, iLOF is computationally expensive. That is why we pre-sample the data before applying it. 

The described process serves two purposes. It discovers if new structures are present in the incoming buffer, considering the existing sample, and it also samples the data so that the projection function can be updated accordingly. The idea is if $\overline{X}_r = \emptyset$ then $f^r = f^{r-1}$, otherwise a new projection function needs to be calculated considering the new sample. Notice that, since we are using an out-of-sample strategy to project the data, the first step is to project $\overline{X}_{[1 \ldots r]}$ when building a new function $f^r$. Given that distance-based projections are invariant to rotation,  every time a new sample is projected, we aligned it with the projection of the previous sample $\overline{X}_{[1 \ldots r-1]}$ using Procrustes analysis~\cite{gower2004procrustes}. In this way, we maintain the spatial coherence of the projection as it evolves, similarly in intent to what has been done with dynamic graphs~\cite{Archambault12, Diehl:2002, diehl}

\section{Technique Analysis and Comparisons}
\label{sec:results}

In this section, we present a series of tests and comparisons to attest the stability and sensibility of the proposed technique and confirm its quality compared to other in-sample, out-of-sample, and incremental/online techniques. In these tests we use datasets with different sizes and dimensionalities, enabling the analysis of different scenarios. The first dataset, \textbf{shuttle}, is composed by instances representing log information. The \textbf{mammals} dataset is an artificially generated dataset representing different features of mammals of four distinct classes (dogs, cats, horses, and giraffes). The \textbf{corel} dataset is composed of images of the Corel image collection represented by color histograms. The \textbf{viscontest} dataset corresponds to a sample of one time step of a simulation obtained from the \textit{IEEE Visualization 2008 Contest dataset}. The \textbf{quantum} dataset is related to particle physics and was obtained from the \textit{KDD-Cup 2004}. Finally, the \textbf{fibers} dataset is composed of instances representing fiber tracks obtained from the \textit{2009 Pittsburgh Brain Competition (PBC) - Brain Connectivity Challenge}. Notice that, since we are comparing our approach against non-incremental techniques as well, we opt to use datasets without time stamp to not bias the process. Table~\ref{tab:datasets} presents the size, dimensionality, and the sources of these datasets. 

\begin{table}[htb]
\caption{Datasets used in the comparisons. From left to right, the columns correspond to the dataset's name, size, dimensionality (number of attributes), and source.}
\label{tab:datasets}
\begin{center}
\begin{tabular}{|l|r|r|l|}
	\hline
	\textbf{Name} & \textbf{Size} & \textbf{Dim} & \textbf{Source}\\
	\hline \hline
	\textbf{shuttle} & 43,500 & 9 & \cite{UCI2013} \\
	\hline
	\textbf{mammals} & 50,000 & 72 & \cite{UCI2013} \\
	\hline
	\textbf{corel} & 68,040 & 32 & \cite{UCI2013} \\
	\hline
	\textbf{viscontest} & 100,000 & 10 & \cite{Whalen2008} \\
	\hline
	\textbf{quantum} & 150,000 & 78 & \cite{KDDCUP04} \\
	\hline
	\textbf{fibers} & 250,000 & 30 & \cite{paulovich2011piece} \\
	\hline
\end{tabular}
\end{center}
\end{table}

\vspace{-1em}
\subsection{Technique Analysis}

\begin{figure}[htb]
\centering
    \includegraphics[width=.9\columnwidth]{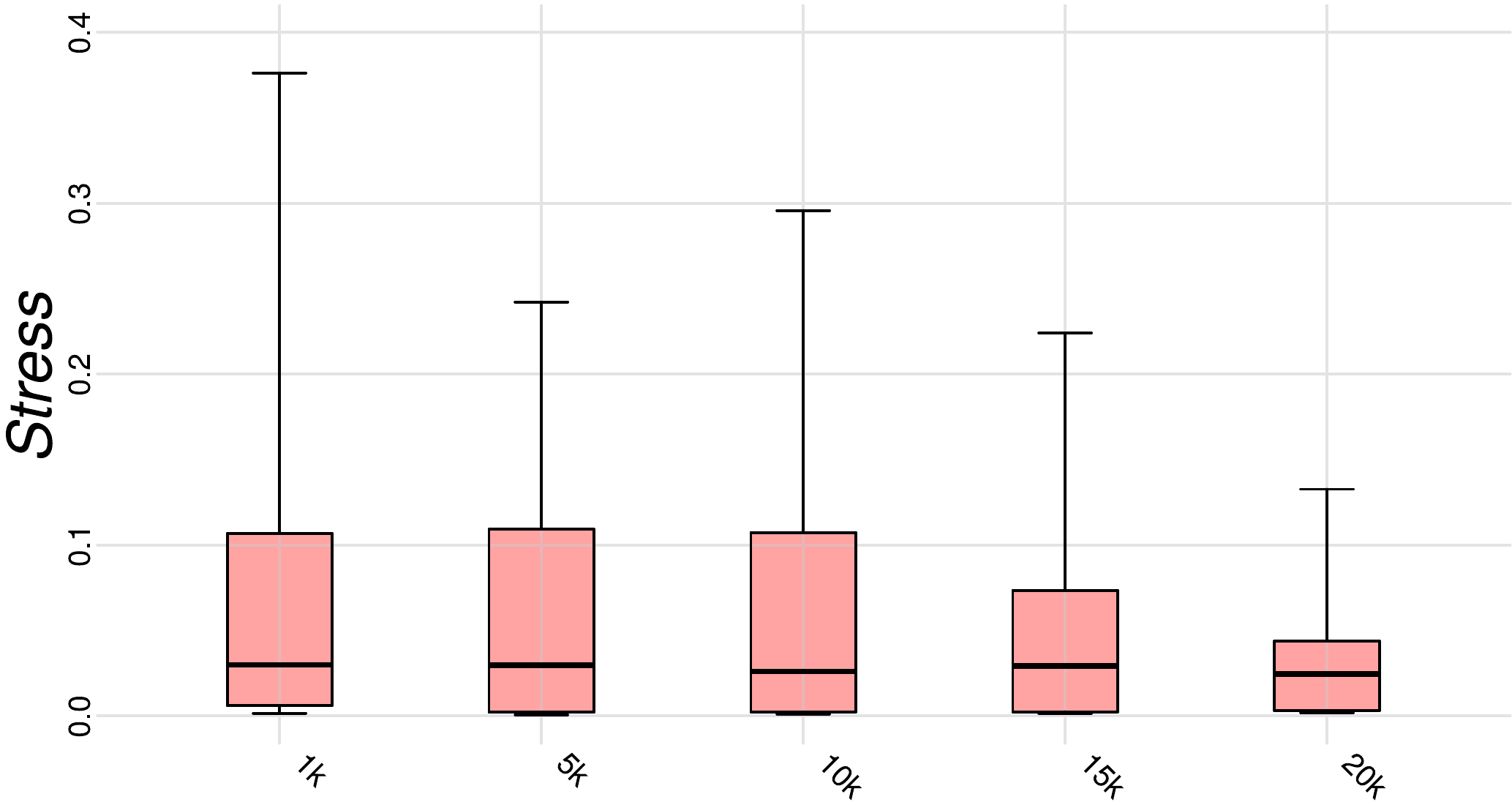}
    \vspace{-0.15cm}
    \caption{Stress boxplot for different buffer sizes. Different buffer sizes produces similar layouts in terms of distance preservation.}
    \label{fig:stress-buffer-size}
\end{figure}

\begin{figure}[htb]
\centering
    \includegraphics[width=.9\columnwidth]{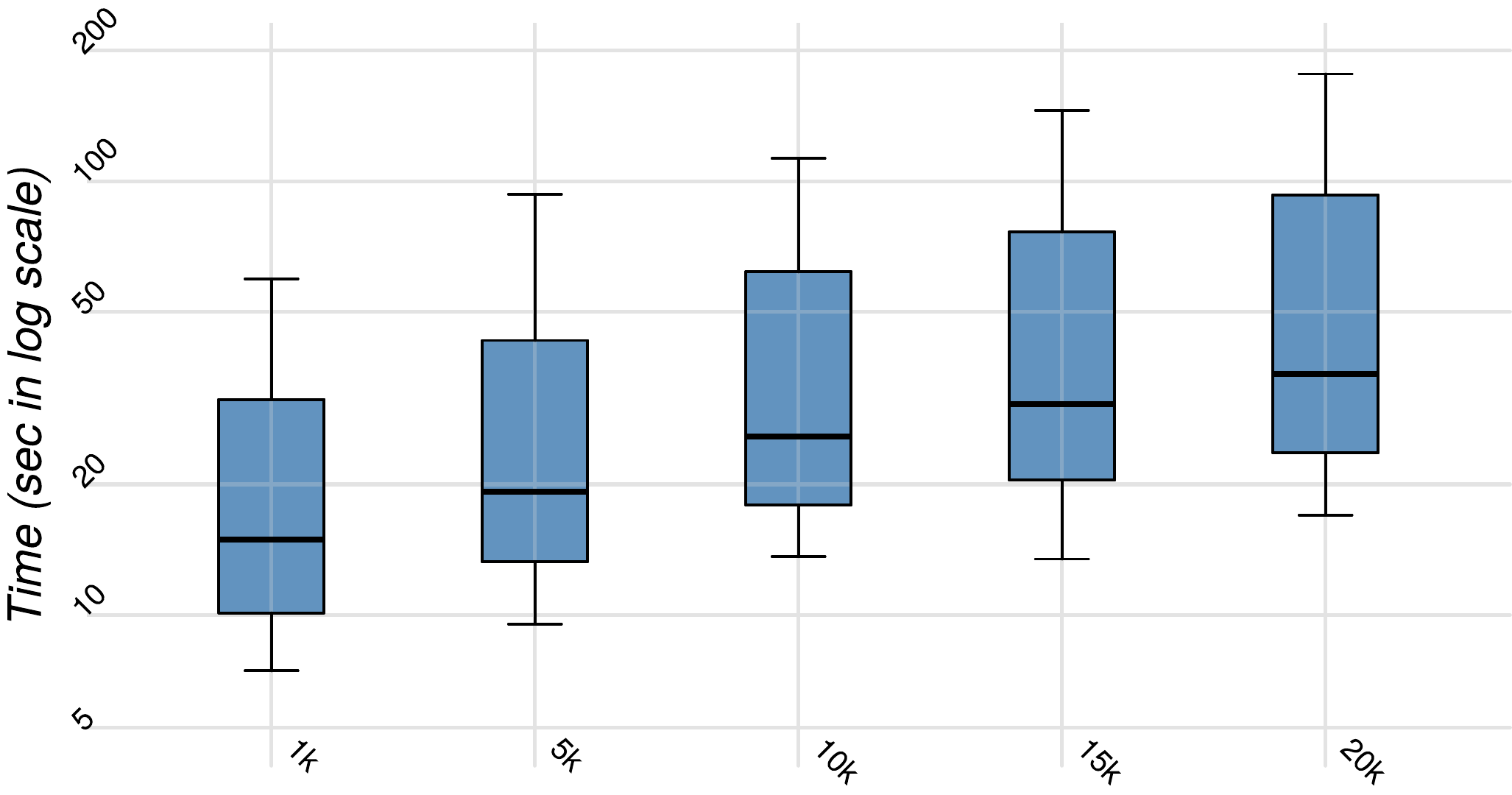}
    \vspace{-0.25cm}
    \caption{Time boxplot for different buffer sizes. The larger the buffer the longer the processing time.}
    \label{fig:time-buffer-size}
\end{figure}

Our first analysis aims to verify the influence of the buffer size on the quality of the produced layout and processing time. In this test, we split each dataset into buffers of varying sizes and present them subsequently to the \textit\techabbr~technique as if it is streaming data. The buffer sizes tested were: $1,000$, $5,000$, $10,000$, $15,000$ and $20,000$. To measure the quality, we use the \emph{normalized stress function}, given by $\sqrt{{\sum_{i<j}{(\delta(x_i,x_j)-d(y_i,y_j))}^2}/{\sum_{i<j}{d(x_i,x_j)^2}}}$, which range in $[0,1]$ with smaller values representing better results. We use stress in our tests because the goal of our approach is to preserve global distance relationships, which is typically the case for huge datasets where fine-grained details are less important than the overall picture. Figure~\ref{fig:stress-buffer-size} shows the results for the datasets of Table~\ref{tab:datasets}. Irrespective of the buffer size, the stress presents similar values on average, with marginally better results for the largest buffer. In terms of running times, Figure~\ref{fig:time-buffer-size} shows that the larger the buffer, the longer the running time, which is expected due to the cost of the sampling and change detection step. Based on that, in the rest of this text, we set the buffer size to $1,000$ instances.

\begin{figure}[htb]
\centering
    \includegraphics[width=.9\columnwidth]{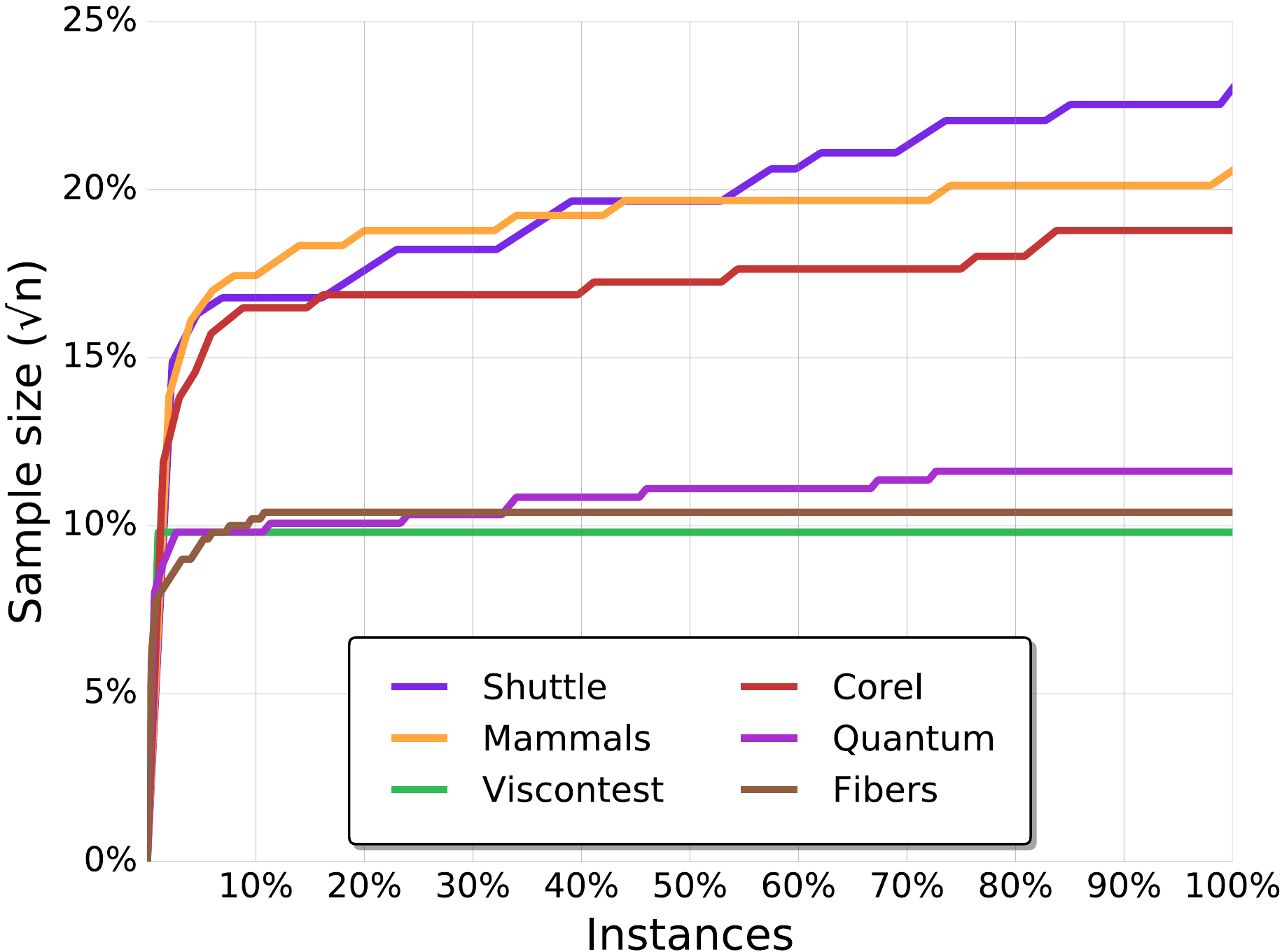}
    \vspace{-0.15cm}
    \caption{Sample size vs. received buffers. For the tested datasets, our cluster-based change detection mechanism constructs a sample never larger than a typical upper-limit adopted by most out-of-sample techniques ($\sqrt{n}$), and normally stabilises it after receiving the first partitions.}
    \label{fig:sampleSize}
\end{figure}

The second analysis aims at verifying if the change detection mechanism produces samples of reasonable sizes, and if the sampling stabilizes as the buffers are received. Figure~\ref{fig:sampleSize} presents the results for the datasets of Table~\ref{tab:datasets}. The vertical axis represents the percentage of the upper-bound limit ($\sqrt{n}$) for the sample size typically employed by state-of-art out-of-sample projection techniques~\cite{Joia2011,paulovich2010plmp}. The horizontal axis represents the percentage of dataset received by the technique (buffers with $1,000$ instances). Notice that the sample sizes never hit the sample upper-bound for the tested datasets. Actually, for most of them, it stays well below and stabilizes after receiving few buffers, showing that our change detection mechanism, although very simple, can successfully distinguish and capture new information as the streaming projection process is executed. 


\begin{figure}[htb]
\centering
    \includegraphics[width=.9\columnwidth]{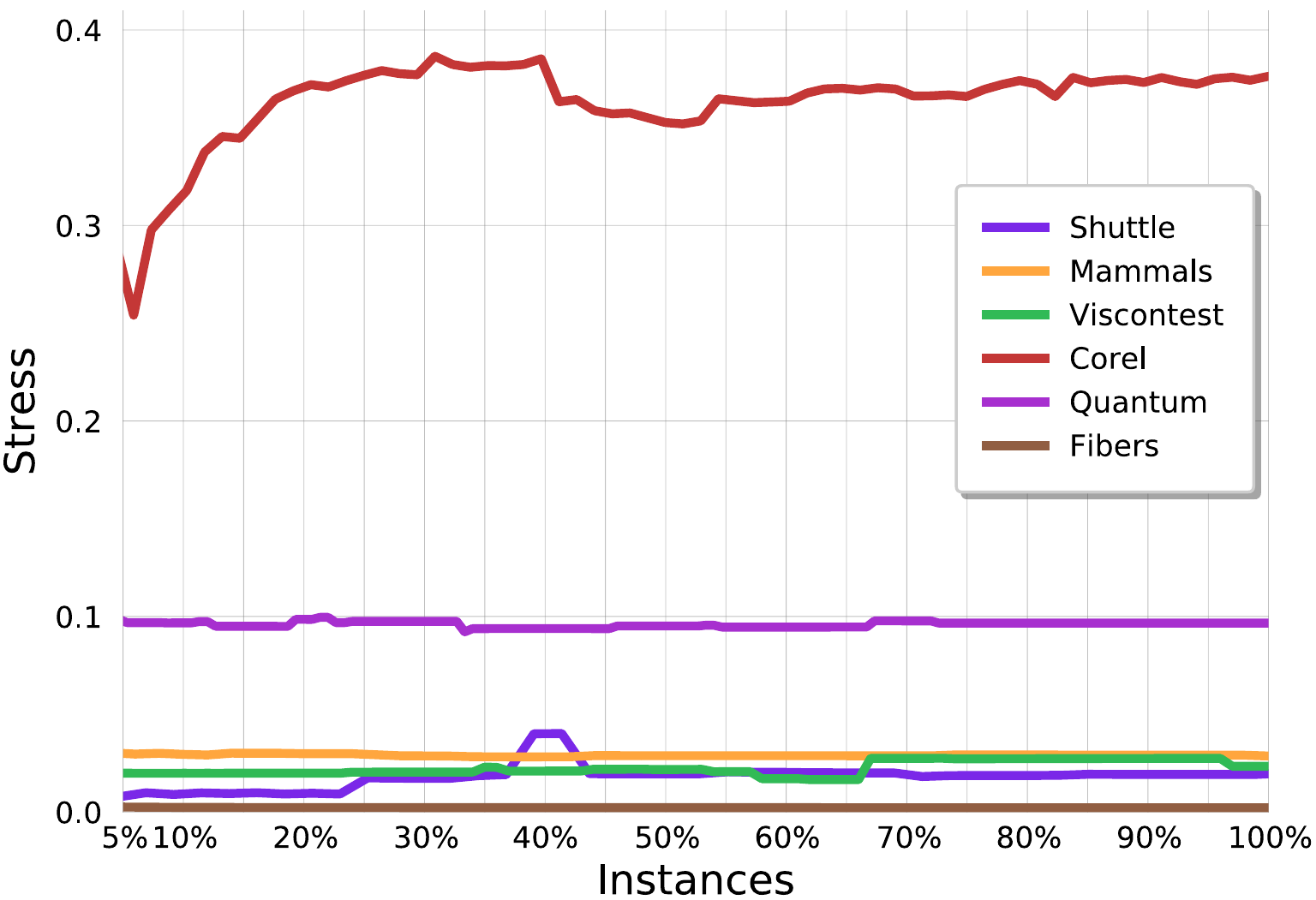}
    \vspace{-0.15cm}
    \caption{Stress vs. received partitions. For most datasets, the stress does not increase over the time, so the re-projection strategy is a good approximation and does not negatively impact the quality of the produced projection.}
    \label{fig:stressEvolution}
\end{figure}

Next, we seek to verify if the re-projection mechanism (see Section~\ref{sec:reproject}) impair the quality of the produced layout in terms of distance preservation as the partitions are processed. As detailed before, every time a new buffer is received, the previous buffers are adjusted to the new information. However, since no data is stored, the multidimensional distances are approximated, replacing them by the two-dimensional distances to re-project the already projected instances. In this test, we measure the projection quality as each buffer is received. Figure~\ref{fig:stressEvolution} shows the results for the datasets of Table~\ref{tab:datasets}. For most datasets, the stress presents similar values from the beginning to the end of the process or early stabilizes as the buffers are processed. Only the \textbf{corel} dataset presents variation since it is a difficult dataset to handle, and all techniques we compare with have problems to produce a precise projection. This gives evidence that the re-projection strategy is a good approximation and does not negatively affect the quality of the final produced projection. 

\begin{figure}[htb]
\centering
    \includegraphics[width=.9\columnwidth]{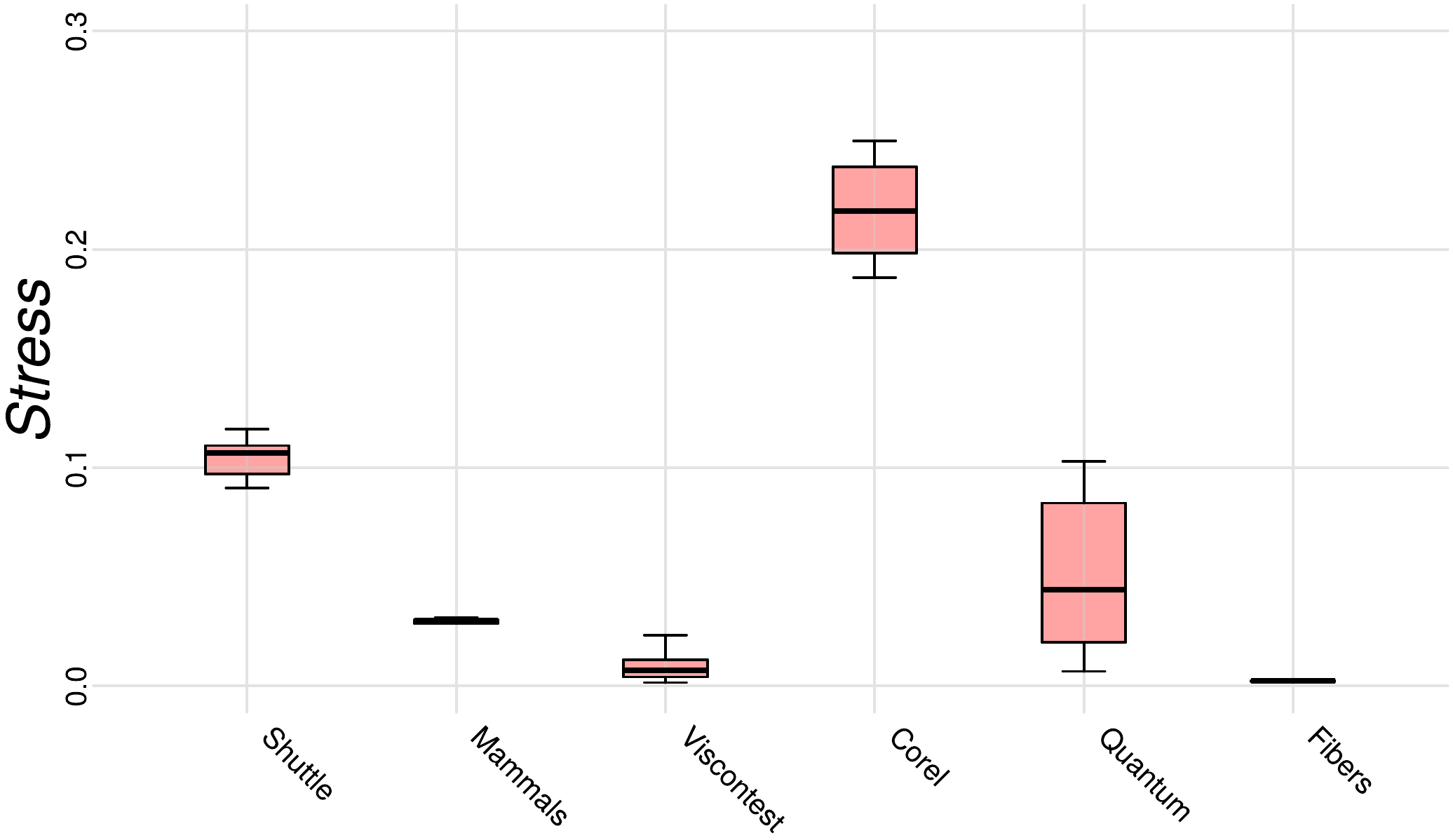}
    \vspace{-0.25cm}
    \caption{Stress boxplots using \textit\techabbr~randomly shuffling the datasets multiple times to change the order the data instances are processes. The small stress variation for each dataset indicates that \textit\techabbr~produces consistent results on different execution scenarios, assuring a good degree of stability and reproducibility.}
    \label{fig:stress-xtreaming-datasets}
\end{figure}

One final analysis was conducted to verify if the technique is sensitive to data ordering. In this test, \textit\techabbr~is executed $30$ times for each dataset, randomly shuffling the order the data instances are presented to the technique (the data instances are shuffled before the data is processed). Again, we split each dataset into buffers and present them subsequently to \textit\techabbr~as streaming data. Figure~\ref{fig:stress-xtreaming-datasets} presents boxplots summarizing the attained results in terms of distance preservation. Observe that the stress variation for each dataset is small, indicating that the overall quality of the produced layouts is independent of the order the data is processed. This is an essential feature for a data streaming technique since the results are consistent in different execution scenarios, assuring a reasonable degree of stability and reproducibility.

In summary, with these tests, we show that \textit\techabbr~technique is not affected by the buffer size presenting similar results in terms of global distance preservation as the buffer varies. Also, the projection quality is not affected by the re-projection phase or by the change detection and sampling strategy we use. The same level of distance preservation is kept from the beginning to the end of the process. Finally, our technique is not sensitive to data ordering showing a good degree of stability and reproducibility.

\vspace{-1em}
\subsection{Comparisons}

In this section, we present a set of comparisons involving batch and incremental projection techniques, checking the global distance preservation. We start by comparing \textit\techabbr~against batch techniques. The batch techniques employed in these comparisons were chosen based on two criteria: they have to be designed to handle large datasets, and present a good tradeoff in terms of quality and running time.  We compare \textit\techabbr~against LAMP~\cite{Joia2011}, UPDis~\cite{neves2018updis}, PCA~\cite{jolliffe1986principal}, PLMP (batch version)~\cite{paulovich2010plmp}, LMDS~\cite{tenenbaum00global}, and Pekalska~\cite{pekalska1999new}. All results reported in this section were produced on an Intel\textsuperscript{\textregistered} Core\textsuperscript{TM} i7 CPU 2.40GHz, with an NVIDIA\textsuperscript\textregistered GeForce GTX 765M video card and 16GB of RAM. To provide a fair comparison, all techniques were implemented in Java, including \textit\techabbr.

\begin{figure}[ht]
\centering
    \includegraphics[width=.9\columnwidth]{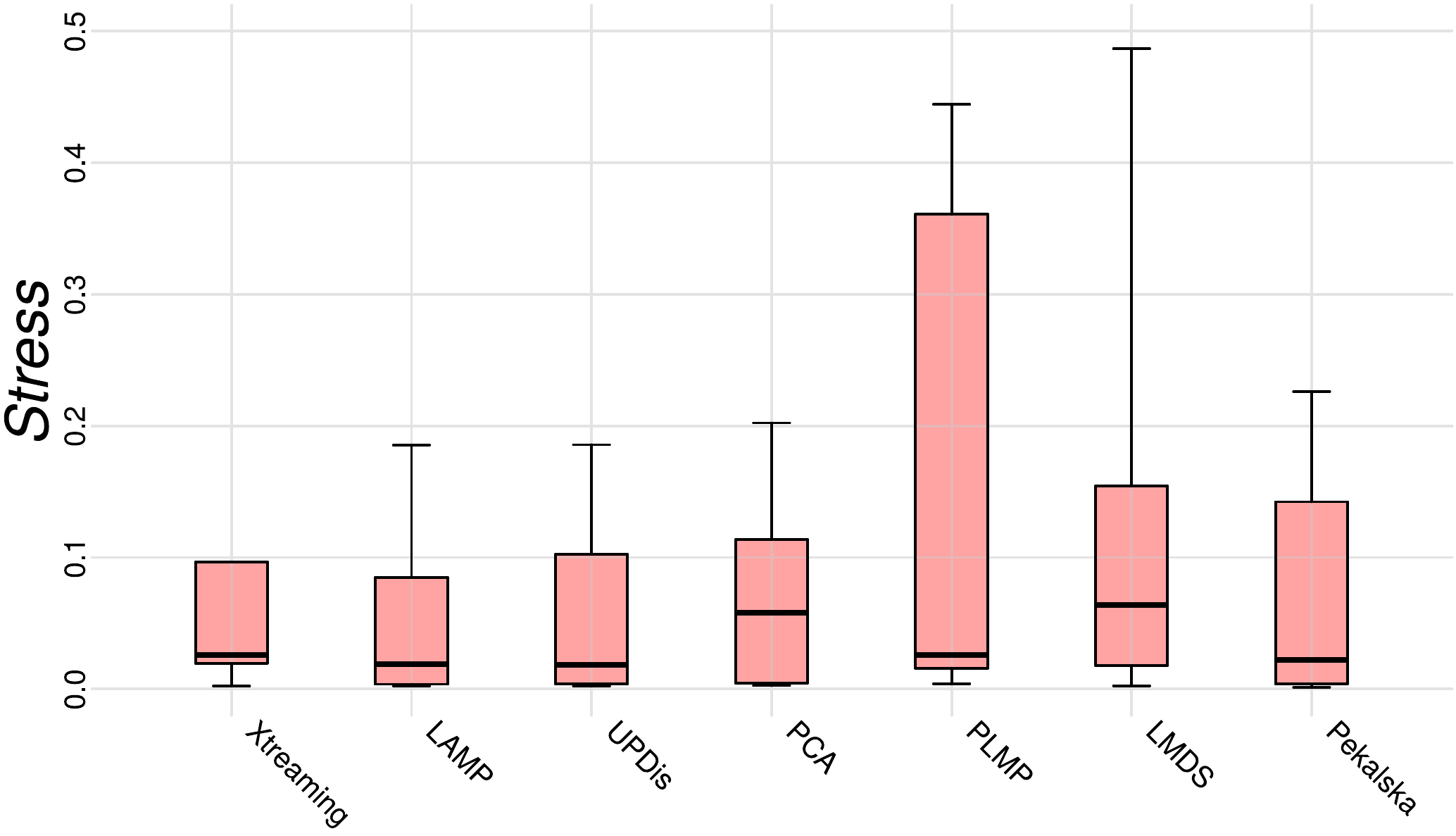}
    \vspace{-0.25cm}
    \caption{Stress boxplots. Comparing to batch (non-streaming) techniques, \textit\techabbr~presents very competitive results, rendering a reliable process to project streaming data.}
    \label{fig:stress-non-stream}
\end{figure}

Figure~\ref{fig:stress-non-stream} presents boxplots summarizing the results in terms of stress. On average, \textit\techabbr~produces  better results if compared to the PCA and LMDS techniques. PLMP produces similar results on average but with a significant deviation on quality, presenting non-reliable results for the tested datasets. Compared to the most precise techniques, LAMP, UPDis, and Pekalska, \textit\techabbr~produces very competitive results. This is an unforeseen outcome since \textit\techabbr~has an additional step, the re-projection, which approximates the multidimensional distances replacing them by the bi-dimensional distances (see Section~\ref{sec:reproject}). Especially, if compared to UPDis, we expected to produce considerably worse results since this is the technique we use to construct the projection functions (see Section~\ref{sec:projfunc}). In practice, however, UPDis is only slightly better. Besides the re-projection, another difference between \textit\techabbr~and the out-of-sample techniques is the mechanism employed to select the sample instances. For all out-of-sample techniques, we use a clustering technique to produce $\sqrt{n}$ clusters, getting their medoids as samples. For \textit\techabbr, the sample is constructed as the data is received, without fixing the number of samples, albeit the sample never gets larger than $\sqrt{n}$ (see Figure~\ref{fig:sampleSize}).

\begin{figure}[!h]
\centering
\includegraphics[width=.9\columnwidth]{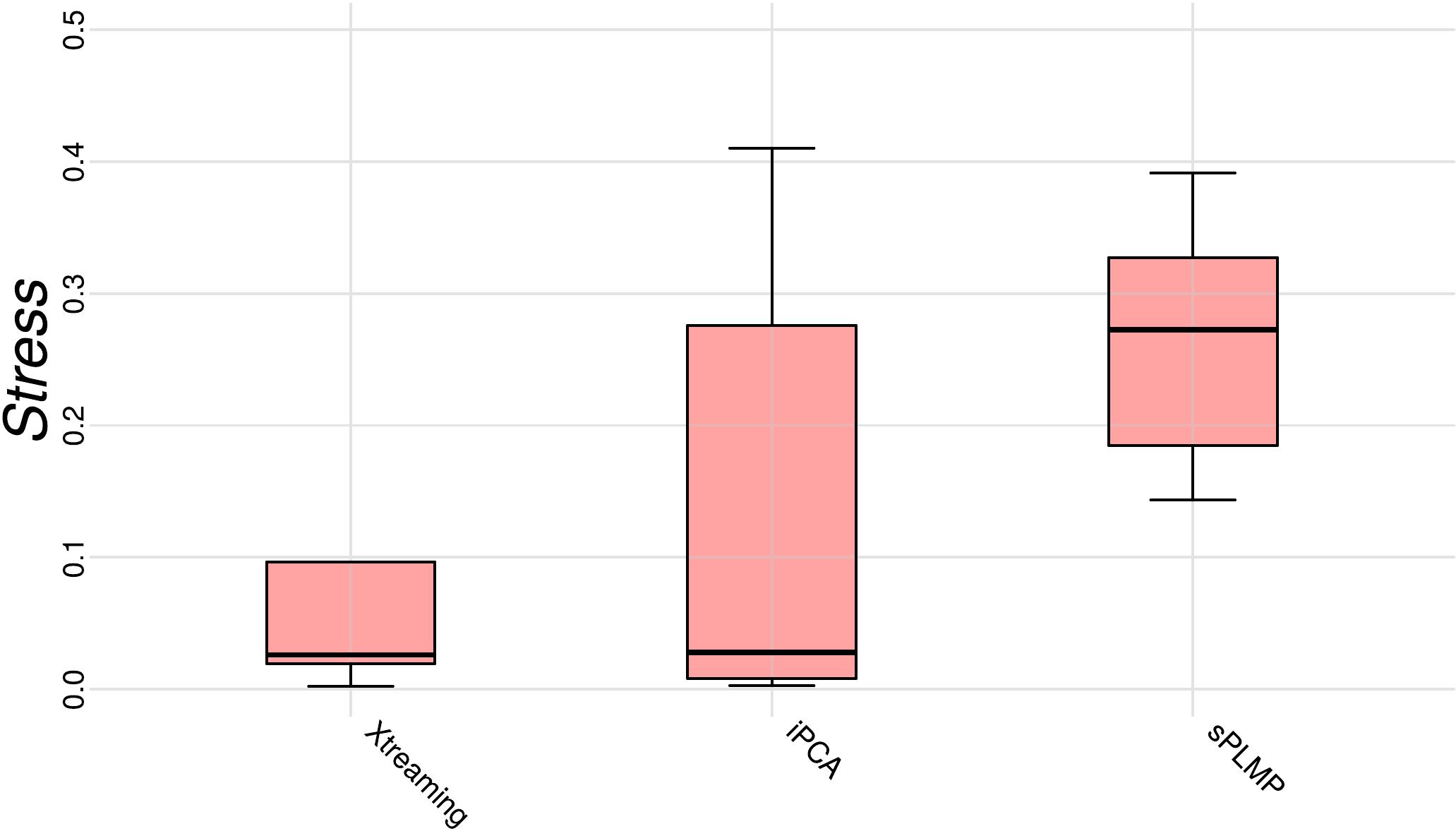}\label{fig:stress-streaming}
\includegraphics[width=.9\columnwidth]{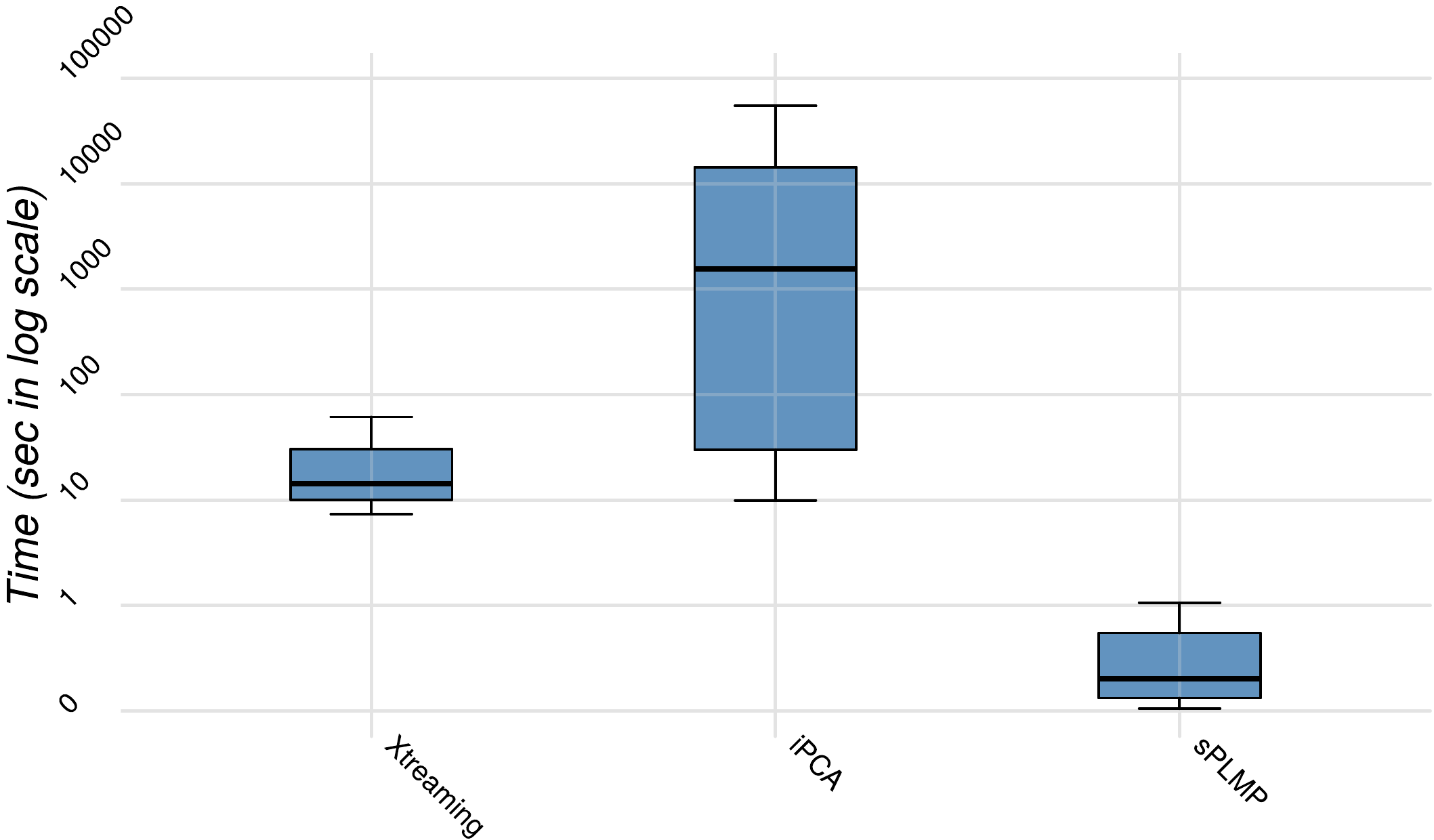}\label{fig:times-streaming}
\vspace{-0.25cm}
\caption{Stress and running times boxplots of the incremental (streaming) projection techniques. \textit\techabbr~presents considerably better global distance preservation than sPLMP, and similar results, on average, to iPCA. However, it is two orders of magnitude faster than iPCA, effectively being a good candidate to process streaming data in real-time.}\label{fig:boxplots-streaming}
\end{figure}

In terms of comparing \textit\techabbr~against other incremental (streaming) projection techniques, we conduct two different analyses: distance preservation and running times. Figure~\ref{fig:boxplots-streaming} presents boxplots summarizing the results. \textit\techabbr~produces considerably better results when compared to the streaming version of the PLMP (here called sPLMP). This is an expected outcome since sPLMP builds a projection model without considering the structures present in the data and use it to project the entire streaming without changing or adapting it over time. Compared to incremental PCA~\cite{Ross2008} (iPCA),  \textit\techabbr~presents very similar results on average but with less deviation in the overall distance preservation, being a more stable streaming technique. It is worth noticing that, when comparing iPCA against PCA (Figure~\ref{fig:stress-non-stream}) it is possible to observe that the iPCA strategy of incrementally updating the projection function introduces errors to the projection process while the same drop in quality is not perceived when comparing the  \textit\techabbr~against UPDis (Figure~\ref{fig:stress-non-stream}). 

In terms of running time, \textit\techabbr~and iPCA are slower than sPLMP. This is expected since sPLMP does not update the projection function over time, and the projection process is reduced to a matrix-vector multiplication. However, the overall distance preservation is penalized, presenting unreliable and low-quality stress.  As mentioned before, \textit\techabbr~and iPCA present similar quality, but \textit\techabbr~is two orders of magnitude faster than iPCA. This happens because the process employed by iPCA to update the eigenvectors is expensive, and it is executed every time a new data instance is received (the nature of iPCA is to be incremental, not streaming). Thereby, \textit\techabbr~presents a better balance between quality and running times, effectively allowing real-time progressive analysis to take place, something that is not supported by the other techniques and an essential feature for the current big data scenarios.

\begin{figure*}[!htb]
    \centering
    \includegraphics[width=.9\textwidth]{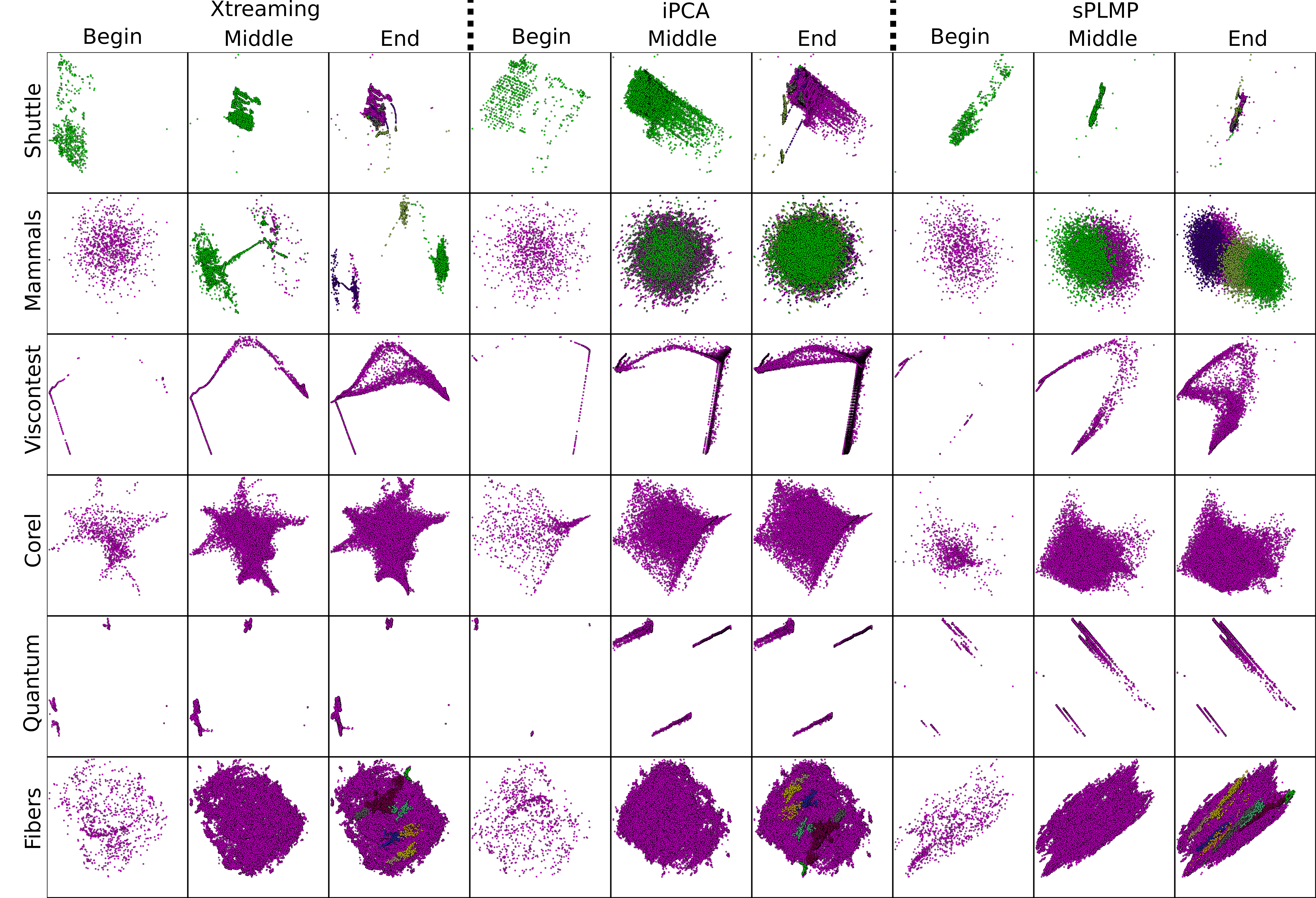}
    \caption{Projections generated by \textit\techabbr, iPCA, and sPLMP techniques at three different stages of the streaming projection process, begin, middle, and end. The datasets are ordered based on the class so that the first instances belong to the same class, and instances of other classes are added to the projection process subsequently. In general, the sPLMP projections are of low quality and are not reliable. Although iPCA projections are of better quality, when the projection process starts with instances that do not represent the overall distribution of the dataset, the technique is not able to distinguish the emerging patterns from the existing ones (\textbf{mammals} dataset). This issue is not observed in \textit\techabbr~projections, and it produces more coherent layouts, successfully capturing the global changes that occur over time.}
    \label{fig:streaming-techs}
\end{figure*}

Finally, Figure~\ref{fig:streaming-techs} presents projections produced by \textit\techabbr, iPCA, and sPLMP techniques at three different stages of the projection process, begin, middle, and end. Begin represents $33\%$ of the dataset, middle $66\%$, and end $100\%$. In this test, we intentionally order the datasets based on the class so that the first instances belong to the same class, and instances of other classes are added to the projection process subsequently. The idea is to show the impact of not updating the data instances already projected as new structures or patterns, in this case, classes of instances, emerge over time. Notice that most of the datasets we used do not present classes, or the notion of class is not clearly defined in terms of distance relationships, that is, in terms of compact and well-separated clusters of similar instances. Only the \textbf{mammals} dataset contains (four) different classes. As discussed (see Section~\ref{sec:related}), sPLMP does not suffer from this issue, it is order-independent, and the final mammals' projection presents the four distinct classes. However, in general, the projections are of low quality (see Figure~\ref{fig:boxplots-streaming}), and are not reliable. The iPCA technique produces the most interesting result. Although, on average, iPCA projections are of good quality (see Figure~\ref{fig:boxplots-streaming}), when the projection process starts with instances that do not represent the overall distribution of the dataset, the technique is not able to distinguish the emerging patterns from the existing ones, even if they are very different. The result is producing visual artifacts where such patterns overlap, in this case, the four different classes of the \textbf{mammals} dataset. This problem is not observed in \textit\techabbr~projections, attesting that the re-projection process is effective even when the dataset distribution changes, producing more coherent layouts that represent well the global structure of multidimensional datasets; a crucial aspect for streaming scenarios where the data is not known before starting the process and can arbitrarily change over time.



\section{Case Studies}

In this section, we describe two case studies that showcase the suitability of using \textit\techabbr~to support the visual analysis of time-varying multidimensional datasets, the first with an artificially-designed dataset, and the second with sentiment and stance analysis on data from Twitter.

\vspace{-1em}
\subsection{Case Study 1: Time-varying Clusters} \label{sec:cs1}

\begin{figure*}[htb]
    \centering
    \includegraphics[width=\textwidth]{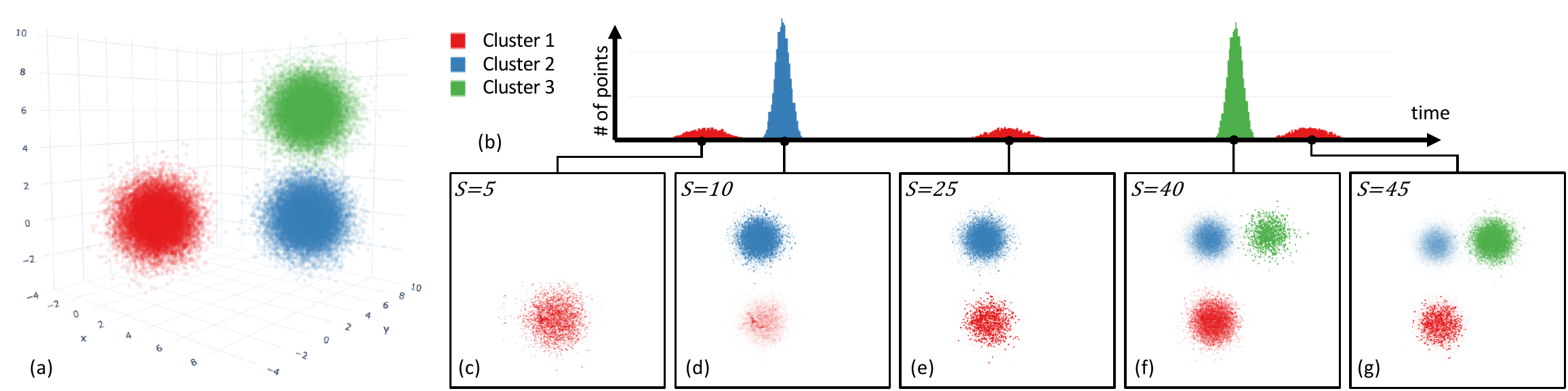}
    \caption{A demonstration of the stability of time-varying patterns in \textit\techabbr 's projection results. The artificially-generated dataset consists on $50,000$ points located in three 3D clusters (a) arriving in the stream at different times (b). As it can be seen both in partial projections (c-f) and in the final layout (g), \textit\techabbr-maintained the overall structure of the dataset, both locally (within clusters) and globally (between clusters), even if the seasonal patterns of Cluster 1, and the two neighbor Clusters 2 and 3, do not overlap in time.}
    \label{fig:clusters}
    \vspace{-1em}
\end{figure*}

In order to support the visual analysis of multidimensional data streams, three requirements are desired from a projection technique that focuses on distance preservation:

\begin{itemize}
    \item \textbf{R1:} It must always maintain a consistent view of the dataset as a whole, even if different patterns happen at different time steps. This means that not only the patterns themselves must be consistent, but the inter-pattern relationships must stay consistent (e.g., distances between clusters).
    \item \textbf{R2:} It must fulfill \textbf{R1} without becoming too ``rigid'' as time passes, so that new patterns may still correctly find their way into the projection.
    \item \textbf{R3:} It must ``learn'' patterns that happen seasonally, so that different occurrences of the same pattern, even if separated by many time steps, will still be correctly identified as being the same.
\end{itemize}

This first case study (Figure~\ref{fig:clusters}) is designed to demonstrate that \textit\techabbr~fulfills these three requirements, and to provide a solid foundation for trusting the visual results of the technique when dealing with streams that contain multiple patterns that happen at different time steps, with little to no overlap between each other. It is based on an artificially-generated dataset with $50,000$ points spread evenly among three 3D clusters (Figure~\ref{fig:clusters}(a)), arriving in the stream according to a fourth (time) dimension (Figure~\ref{fig:clusters}(b)). Both the spatial and the time dimensions of the dataset were designed carefully to test the three requirements stated previously. The centroids of the clusters are positioned in three different vertices of an invisible cube, such that Clusters 2 and 3 (\emph{blue} and \emph{green} respectively in Figure~\ref{fig:clusters}) are in neighboring vertices, so they should be positioned closer to each other, while Cluster 1 (\emph{red} points in Figure~\ref{fig:clusters}) is in a diagonally-opposite vertex. Regarding time, the points from Cluster 1 are evenly distributed in three different non-overlapping occurrences (simulating a seasonal pattern), while the points from Clusters 2 and 3 are very well-separated in time, also with no overlaps.

The dataset is analyzed by \textit\techabbr~as a stream of $50$ consecutive steps ($S$) of $1,000$ points each. Figures~\ref{fig:clusters}(c-g) show partial projections of the stream in key time steps. In order to denote the ``age'' of a point $p$, i.e. in which step $S_p$ it arrived, we map its opacity $O_p$ between $0$ (fully transparent) and $1$ (fully visible) according to the transfer function
\begin{equation} \label{eq:op}
O_p=\frac{1}{S-S_p+1},
\end{equation}
so that, points from the current step ($S$) are always fully visible.

At $S=5$ (Figure~\ref{fig:clusters}(c)), we still see only Cluster 1, with no sign of any other pattern. Cluster 2 can be clearly seen at $S=10$, and while no new points of Cluster 1 have arrived in the stream for a while, they can still be seen with low opacity $O_p$. At this point it is interesting to notice that Cluster 2 has been positioned correctly regarding its separation to Cluster 1, even though more than 5,000 data points had already been analyzed before the points from Cluster 2 started to arrive. This is related to both \textbf{R1} (global consistency \emph{between} clusters) and \textbf{R2} (learning new patterns). 

At $S=25$ (Figure~\ref{fig:clusters}(e)), Cluster 1 has returned to the stream after a long period of inactivity, and its points were correctly positioned on top of the previous points from the same cluster, showing that \textit\techabbr~has been able to maintain its \emph{memory} about previous patterns in a lightweight way. Some other techniques that might use \textit\techabbr~as a pre-processing step for learning from the stream, for example, would be able at this point to correctly identify that the new points belong to the same cluster, even though all clusters had stopped providing points a long time ago. This is mostly related to \textbf{R3} (learning seasonal patterns), but also to \textbf{R1} (global consistency \emph{within} clusters).

Points from Cluster 3 have arrived and settled in the projection by $S=40$ (Figure~\ref{fig:clusters}(f)). Notice that, by the time its points started to arrive, more than $35,000 points$ had already been analyzed, with two occurrences of Cluster 1 and one occurrence of Cluster 2. Nevertheless, the projection did not have problems to learn the new pattern and position the cluster correctly, not only well-separated from the other clusters but positioned closer to its neighbor (Cluster 2) and far from Cluster 1. This is the expected outcome based on how the cluster centroids were initially generated in different vertices of the cube. These observations are related to \textbf{R1} (global consistency \emph{between} clusters) and \textbf{R2} (learning new patterns). 
Finally, at $S=45$ (Figure~\ref{fig:clusters}(g)), more points from Cluster 1 arrive at the stream, again being correctly positioned on top of their previous intra-cluster neighbors, providing more evidence for \textbf{R3} (learning seasonal patterns) and \textbf{R1} (global consistency \emph{within} clusters). 

\begin{figure*}[htb]
    \centering
    \includegraphics[width=\textwidth]{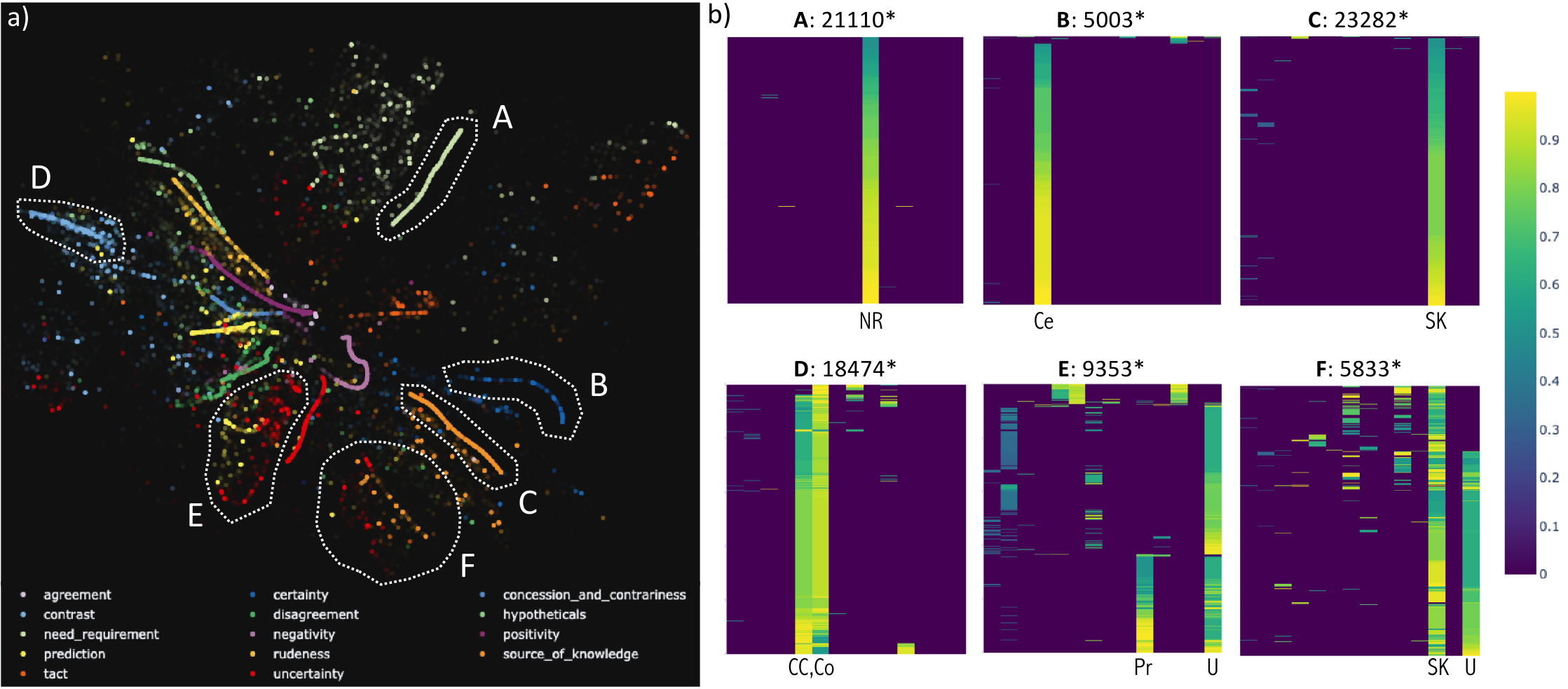}
    \caption{A case study on using \textit\techabbr~for the visual analysis of a social media stream from Twitter characterized by sentiment and stance categories. Each point is a sentence from a tweet, the colors indicate the strongest detected sentiment/stance in that tweet, and the opacity indicates how ``old'' it is. At first, some patterns can be readily detected, such as groups of tweets with predominantly (A) \emph{need and requirement}, (B) \emph{certainty}, or (C) \emph{source of knowledge}. More investigation shows that even the less-identifiable groups can be traced to consistent patterns in feature interactions, such as (D) \emph{contrast} vs \emph{concession and contrariness}, (E) \emph{prediction} vs \emph{uncertainty}, and (F) \emph{source of knowledge} and \emph{uncertainty} (see the text for more details on these groups).}
    \label{fig:twitter}
    \vspace{-1em}
\end{figure*}

One final general observation about \textit\techabbr~that can be recovered from this case study as a whole is the consistency of its output throughout time steps. Consecutive projections of different time steps of the dataset maintained the overall shape of the data very closely at all times, even after long periods of inactivity and little overlap between the time distributions of different clusters. This is also a critical desirable characteristic of a technique that intends to support visual analysis of time-varying data.

\vspace{-1em}
\subsection{Case Study 2: Sentiment/Stance Analysis on Twitter}

\paragraph*{Data set description.}

The data set consists of social media posts in English collected from the Twitter streaming API during 12:00:00--23:59:59 UTC on July 9, 2018.
The keywords used to filter the stream of tweets include several phrases and names related to Brexit,
such as ``brexit'', ``british referendum'', ``bnp'', ``ukip'', and ``nigel farage''.
The particular date chosen for this data set corresponds to the day of resignation of the UK Foreign Secretary Boris Johnson,
which was widely discussed on both traditional and social media.
The texts of the tweets are split into individual sentences (utterances), which are classified according to sentiment~\cite{Mohammad2016_survey} and stance~\cite{Mohammad2017}.
Sentiment classification is carried out with the VADER classifier~\cite{Hutto2014} that reports normalized output weights for the categories \emph{positivity} and \emph{negativity}.
Weights below a certain threshold (in our case, $0.3$) are ignored and such utterances are considered neutral (they assume the value $0.0$).
Stance classification is carried out with a custom stance classifier \textbf{[anon. ref.]} that detects multiple categories and reports the corresponding output weights (the values of the weights range within the interval $[0.5, 1.0]$, anything less than $0.5$ is reduced to $0.0$).
The exact set of 12 stance categories 
includes the following: \emph{agreement} (A), \emph{certainty} (Ce), \emph{concession and contrariness} (CC), \emph{contrast} (Co), \emph{disagreement} (D), \emph{hypotheticals} (H), \emph{need and requirement} (NR), \emph{prediction} (P), \emph{rudeness} (R), \emph{source of knowledge} (SK), \emph{tact} (T), and \emph{uncertainty} (U).
In its final form, the dataset includes $503,511$ rows (sentences extracted from tweets) and $14$ columns with the classification results (two columns for sentiment and twelve for stance) represented as numbers in the range $[0.0; 1.0]$.

The final projection of the full data set, generated by \textit\techabbr, is shown in Figure~\ref{fig:twitter} (a). The colors indicate the strongest detected sentiment/stance in each data point, or in other words, the sentiment/stance label with the maximum value of the corresponding row. They are provided as support to the visual analysis, but should not be considered as the final classification of the data points. The opacity, as in Section~\ref{sec:cs1}, indicates how ``old'' each data point is, following Equation~\ref{eq:op}. In order to investigate in more detail different groups of points, a heatmap is generated from a selection by mapping the values of the features into a continuous colormap ranging from $0.0$ to $1.0$. Each column of the heatmap (the $x$ axis) represents one of the $14$ features, and the $y$ axis represents the selected points. The number of selected points is shown on top of each heatmap, but if a selection includes more than $5,000$ points (marked with an asterisk), they are downsampled to this number, to maintain efficiency. Additionally, points are arranged in the $y$ axis according to a PCA projection down to $1$ dimension, so that similar points are grouped together.

At first, some patterns/groups can be readily detected due to \textit\techabbr~having isolated them clearly from others, such as tweets with predominantly \emph{need and requirement} (Group A), \emph{certainty} (Group B), or \emph{source of knowledge} (Group C), as shown by the homogeneous feature distributions in the corresponding heatmaps. This initial step of detecting and understanding the easily-identifiable groups can be associated with an \emph{overview} task of the projection, which \textit\techabbr~supports for this dataset. 
After the overview, the analyst might be interested in exploring whether the less-identifiable groups--those that are less clearly separable or salient in the layout--are also meaningful, or are simply the results of noise in the data or lack of accuracy in \textit\techabbr. 
A more involved investigation of specific areas of the layout shows, however, that even these less-identifiable groups can be traced to consistent patterns in feature interactions, as described next.


\paragraph*{Group D: Contrast vs. Concession and Contrariness.}
Group D initially looks like a homogeneous group of points detected as mostly \emph{contrast}, but the heatmap shows an almost-equal distribution of \emph{concession and contrariness}. 
These are related categories, but the former does not include instances of \emph{antithesis} as studied in rhetorical structure theory~\cite{Azar1999} (for instance, negations or antonyms).
It is thus expected that these categories would co-occur in the classification results, and the fact that the projection technique has positioned the corresponding utterances in the same neighborhood confirms our expectations.

\paragraph*{Group E: Prediction vs. Uncertainty.}
The co-occurrence of categories such as \emph{prediction} and \emph{uncertainty} was previously observed in multiple data sets by our collaborators in linguistics, and it can be explained by the uncertain nature of most predictive statements (e.g., ``I guess the political crisis will continue\dots'').

\paragraph*{Group F: Source of Knowledge vs. Uncertainty.}
The presence of a cluster with a prominent number of co-occurrences of such stance categories as \emph{uncertainty} and \emph{source of knowledge} can be explained by the attempts to attribute uncertain information or rumors to third-party sources (e.g., ``According to World Press, the event may have occurred\dots'') or to employ hedging while presenting one's opinion (e.g., ``I think I remember reading about their political failures\dots'').

\section{Discussions and Limitations}

One potential problem that may occur is related to the rate the data is captured or produced in time-dependent streaming applications. In these applications, it is vital to produce the projection layout on-the-fly as the data is received. However, if the rate of data production is larger than the processing time imposed by \textit\techabbr, some data may be discarded. One possible solution is to use a double-buffer strategy where parts of the data can be stored while others are processed. This difficulty is more related to technology than to the technique itself. In practice, we expect that, with appropriate hardware and implementation, \textit\techabbr~runs much faster than the results reported in Section~\ref{sec:results}, which is already orders of magnitude faster than other comparable techniques.

One limitation, which is a direct consequence of \textit\techabbr{}'s strength, is that it is not indicated for small-sized problems. Due to the approximations and the strategies employed to allow a single traversal of the data, in general, for small datasets, \textit\techabbr~is slower than most in-sample and out-of-sample projection techniques, and it is not as precise as them. However, this drop in precision is the price to pay for handling streaming data, and approximated solutions are acceptable in order to enable streaming applications~\cite{Gama2012streaming}. Also, our goal is global distance preservation since, for large datasets, an accurate overview (distances among groups of instances) is typically more important than fine details in small neighborhoods. That is why we only evaluate our results using the stress measure. For local quality measures and smaller datasets, we suggest using other more appropriate techniques, such as t-SNE~\cite{van2008visualizing}, ISOMAP~\cite{tenenbaum00global}, or LoCH~\cite{FADEL2015546}.


\section{Conclusions}

In this paper, we propose a novel multidimensional projection technique called \textit\techabbr, which is shown to be one of the first reliable approaches for projecting streaming data applications. \textit\techabbr~presents different novel strategies, enabling the processing of data as it is received and to adapt the visual layout to new emerging structures without the need for visiting the multidimensional data more than once. The set of comparisons we provide shows that \textit\techabbr~is comparable to existing out-of-sample techniques in terms of global distance preservation while presents a much better trade-off between quality and running time if compared to other online or incremental approaches. Moreover, the potential use of \textit\techabbr~in streaming data scenarios opens possibilities for novel applications to properly process ever-growing data collections, such as the real-time analysis of social media networks. We are currently investigating new visual metaphors for streaming data projections towards giving visual insights about structural changes and trends that emerge over time, a challenge for the next years. 










\bibliographystyle{IEEEtran}
\bibliography{referencias}

\end{document}